\documentclass[12pt]{article}
\usepackage{mheck}

\usepackage{longtable}

\usepackage{blkarray}

\usepackage{tikz} 
\usetikzlibrary{matrix}

\theoremstyle{theorem}

\theoremstyle{definition}

\def\Dsl{\,\raise.15ex\hbox{/}\mkern-13.5mu D}

\date{August, 2018}

\title{Theta-problem and the String Swampland
}

\authors{Sergio Cecotti\footnote{e-mail: {\tt cecotti@sissa.it}}${}^\S$ and Cumrun Vafa\footnote{e-mail: {\tt vafa@g.harvard.edu}}${}^\ddagger$\vskip 9pt

\centerline{${}^\S$ SISSA, via Bonomea 265, I-34100 Trieste, ITALY}
\centerline{${}^\ddagger$ Department of Physics, Harvard University, Cambridge, MA 02138, USA}}

\abstract{In the context of $\cn=2$ supergravity without vector multiplets coupled to hypermultiplets, the coupling constant of graviphoton $\tau$ is apriori a free parameter.   Stringy realization of this and using a mathematical conjecture leads to the statement that $j(\tau)\in \R$
so that the $\theta$-angle is $0$ or $\pi$.  We conjecture that for any consistent realization of $\cn=2$ supergravity theories coupled only to hypermultiplets this is the case and the rest belong to the swampland.
This leads to the speculation that the $\theta$-angle for QCD or QED may also be fixed to $0$ for quantum gravitational consistency.}
\begin{document}

\maketitle

\newpage

A swampland conjecture (for a review see \cite{review}) states that,    in $d\geq4$ space-time dimensions, a low energy effective theory, which has a UV completion to a consistent quantum gravity, cannot have  \emph{free} parameters. According to the conjecture,
the couplings in the effective Lagrangian $L$ are either the v.e.v.\! of light scalar fields or are frozen at some special values which are v.e.v.\! of heavy degrees of freedom of the full theory which are not part of the low-energy description. In the second case, the set $\cs$ of quantum consistent values of the effective couplings has zero-measure in the continuous space of semi-classically permitted parameters.  In particular, if we believe there are finitely or countably many compactifications types (taking into account fluxes), the set $\cs$ should be at most countable and arbitrary continuous values are not allowed. The couplings in $\cs$ are  very non-generic; from the low-energy perspective they appear to be highly fine-tuned to values characterized by special physical properties. 
Understanding the set of consistent couplings  $\cs$ is clearly a basic problem of low energy phenomenology. 
\medskip

The simplest and most basic instance of the question is the actual universe we live in. The only massless degrees of freedom are the graviton $g_{\mu\nu}$ and a photon $A_\mu$. 
In the extreme infrared the effective Lagrangian takes the form\footnote{\ For simplicity we have set the cosmological constant $\Lambda$ to zero. }
$$\label{effLag}
L=\sqrt{-g}\left(\frac{1}{2}R-\frac{1}{4e^2}F^2+\frac{\theta}{32\pi^2} F\tilde F\right),\eqno{(1)}
$$
for some gauge coupling $e$ and angle $\theta$.  Even though in the non-gravitational set-up the $\theta$-angle of a $U(1)$ theory in $\R^4$ is unobservable, in the gravitational context the value of $\theta$ matters: e.g.\!  we can consider gravitational backgrounds with non-trivial 2-cycles, with $\int F\wedge F\not=0$.  Also we can consider the macroscopic entropy of extremal black holes which is a non-trivial function of $\theta$.  In particular the number $N$ of states with electric/magnetic charge $(q,p)$ with minimal mass\footnote{\ Note that $(q,p)$ are integers because the photon needs to be associated to a compact $U(1)$ gauge theory for a consistent quantum theory of gravity \cite{banks}.} is expected to be given, to leading order for large charges, by the entropy of the extremal charged dyonic black holes:
$$
N\sim \exp\!\left(\pi \,{|q+\tau p|^2\over \mathrm{Im}\,\tau}\right)\!,
$$
where $\tau$ is given by
$$
\tau\equiv\frac{\theta}{2\pi}+\frac{4\pi i}{e^2}.
$$
Note that the value of $\tau$ depends on the electro-magnetic duality frame we pick and undergoes $SL(2,\Z)$ transformation under such a change of frame which acts on the complexified gauge coupling $\tau$ as
$$
\tau\rightarrow {a\tau  +b\over c\tau +d},
$$
where $ad-bc=1$ and $a,b,c,d \in \Z$. The space of inequivalent effective Lagrangians of the form (1) is the Riemann sphere $\bP^1$ parametrized by the modular invariant $j(\tau)$ \cite{milne}
$$
j(\tau)= q^{-1}+744+\sum_{n\geq1} c_n\,q^n,\quad\text{where } q=e^{2\pi i \tau},\quad c_n\in\Z.\eqno{(2)}
$$ 
In a quantum theory of gravity we expect $\tau$ to be the expectation value of a field at the minimum of its potential. In other words we always have axion and saxion fields whose vev would lead to $\tau$, and if there are no massless scalars around, this means the corresponding fields have been frozen at their minima giving rise to a specific value of $\tau$.
No known semi-classical consistency requirements fix the value of $\tau$.  But we expect the allowed set to be countable and perhaps even finite. 
The problem is to determine the sparse set $\cs\subset \bP^1$ of photon couplings which are consistent in the context of quantum gravity. Stated differently: to which special values are we supposed to fine-tune the fine structure constant $\alpha\equiv (\mathrm{Im}\,\tau)^{-1}$ and the angle $\theta\equiv 2\pi\,\mathrm{Re}\,\tau$ in order to achieve a consistent quantum gravity theory?  In other words what distinguishes the values of $\tau$ belonging to the string landscape as opposed to the swampland?  
\medskip

 From the low-energy perspective it is not clear which values of $\alpha$ and $\theta$ should be considered ``special''. At weak coupling there are two preferred values of the angle $\theta$ at which something special  happens: at $\theta=0$, $\pi$ the theory preserves time-reversal (at the Lagrangian level).
This statement is frame dependent; its $SL(2,\Z)$-invariant version requires  $j(\tau)$ to be real (as can be seen from its expansion in eqn.(2)):
$$
\text{$L$ is $T$-invariant}\quad\Longleftrightarrow\quad j(\tau)\in\R.\eqno{(3)}
$$
For $\tau$ in the usual modular fundamental domain $\cf$, this corresponds to\footnote{\ Here $\sim$ stands for equality up to $SL(2,\Z)$ equivalence.}
$$
\label{kkzaq}
\tau\in \mathcal{R}\equiv \big\{i/\alpha\;\big|\; 0<\alpha\leq 1\big\}\bigcup \partial\cf \ \sim\ \big\{i /\alpha\;\big|\; 0<\alpha\leq 1\big\}\bigcup \big\{ 1/2+i /\alpha\;\big|\; 0<\alpha<2\big\},\eqno{(4)}
$$
and, modulo $SL(2,\Z)$, $\theta=0$ or $\pi$.
For $j(\tau)\in\R$ we have\footnote{\ \label{spei}For $j(\tau)=1728$ we may set $\theta$ to either 0 or $\pi$ by a $SL(2,\Z)$ transformation since $i\sim(1+i)/2$.}
$$
\theta=0\ \Leftrightarrow\ j(\tau)\geq 1728,\qquad \theta=\pi\ \Leftrightarrow\ j(\tau)\leq 1728.
$$

As a first question, we may ask whether quantum consistency is compatible with $T$-invariance of the photon sector, that is, if the consistent set $\cs$ contains some time-reversal preserving coupling, $\cs\cap \mathcal{R}\neq\emptyset$. More optimistically, we may ask
whether 
$$\label{pooq}
\cs\overset{?}{\subset}\mathcal{R},
\eqno{(5)}$$
 i.e.\! if quantum consistency  \emph{implies} $T$-invariance of the Lagrangian (1). 
\medskip

The determination of the quantum set $\cs$ for the real-world effective Lagrangian (1) is a formidable task. Luckily, there is a simpler version of the problem that can be viewed as a toy model for the actual universe. In the long-wavelength limit, the model (1) is indistinguishable from the bosonic sector of pure $\cn=2$ supergravity which in addition to the graviton has a graviphoton partner.  In a general $\cn=2$ supergravity, the gauge couplings and angles are non-trivial functions of the complex scalars in the vector multiplets, but they cannot depend on the hypermultiplet scalars. Coupling additional hypermultiplets to supergravity has no effect on the deep infrared dynamics of the photon sector, and can be ignored for the discussion at hand.  So, in particular, we can study quantum gravity theories with $\cn=2$ which have no vector multiplets. In such theories the graviphoton coupling $\tau$ would be fixed despite having additional massless fields in the theory.  These are easier to construct in string theory (in fact currently we know of no way to construct $\cn=2$ supergravity theories without any massless vector or hypermultiplets).  For example we can consider type IIB compactifications of Calabi-Yau threefolds.  In such a case, if the complex structure of the Calabi-Yau $X$ is rigid, we have no associated vector multiplets
(but, as many hypermultiplets as $1+h^{1,1}$ where $h^{1,1}$ is the number of K\"ahler moduli of the CY).  In such a setup the complex coupling $\tau$ of the graviphoton is a field-independent $c$-number as in eqn.(1), depending on the choice of the rigid Calabi-Yau $X$, which belongs to a set $\cs_{\cn=2}\subset\bP^1$ of $\cn=2$ quantum   consistent  couplings.\footnote{\ More precisely: $\cs_{N=2}=\cup_{h^{1,1}} \cs(h^{1,1})$ where $\cs(h^{1,1})$ is the consistent coupling set in presence of $h^{1,1}+1$ hypers. We expect $\cs(h^{1,1})\neq \emptyset$ only for finitely many $h^{1,1}\in\bN$. } ${\cn =2}$ supersymmetry by itself does not restrict\footnote{\ The \textsc{sugra} model is described by the prepotential $F(X^0)= -\tau (X^0)^2/2$ where (classically) $\tau$ is an arbitrary point in the upper half-plane.} the allowed values of $\tau$, and the set $\cs_{\cn=2}$ is determined purely by quantum-gravitational consistency.  Moreover we expect this set of values of $\tau$ to be finite, because it is believed that there are only a finite number of Calabi-Yau 3-folds and thus we only have a finite (and small) number of rigid CY 3-folds.\medskip

While finding $\cs_{\cn=2}$ is not equivalent to determining the real-world set  $\cs$, the two problems have qualitatively the same flavor. 
\medskip

The set of $\cn=2$ supergravities (with zero cosmological constant) which can be completed to fully consistent quantum gravity contains (and is believed to coincide with) the set of models obtained by compactifying Type IIB superstring on a Calabi-Yau 3-fold which we restrict to be rigid to avoid vector multiplets.  In this case the rank of $H_3(X,\Z)$ is 2 and we can choose an integral basis of 3-cycles  given by 
$$H_3(X,\Z)=\Z\gamma_1\oplus \Z\gamma_2$$
with $\gamma_1\cdot\gamma_2=1$.  The D3 branes wrapped around these cycles can be viewed as electric and magnetically charged states.  Picking such a basis amounts to fixing the electro-magnetic duality frame.  The graviphoton coupling $\tau$ in this frame can be simply computed as\footnote{\ The extra minus sign in the \textsc{rhs} with respect to the elliptic curve case is due to the fact that $\Omega$ is a 3-form instead of a 1-form.}
$$\tau =-{\int_{\gamma_2}\Omega \over \int_{\gamma_1}\Omega},$$
where $\Omega$ is the holomorphic $(3,0)$ form.

More than 50 examples of (non-isomorphic) rigid Calabi-Yau 3-folds are known \cite{meyer, arithmetics,modular,modular2011}; the total number of rigid CY 3-fold is expected to be finite, and the known ones are presumably a substantial portion of all such manifolds. As algebraic varieties, rigid CY 3-folds are defined over finite extensions\footnote{\ This would follows e.g.\! from the Hodge conjecture \cite{moore2}.}  $\bK$ of $\bQ$; the rigid CY 3-folds defined over $\bK$ organize in orbits of the Galois group.\footnote{\ Number theoretical aspects of $\tau$ related to $\cn=2$ supergravity have been discussed in ref.\cite{moore1,moore2}.} Since rigid CY are very rare objects, the orbits should be short, i.e.\! the degree of $\bK$ small.
Indeed almost all known examples correspond to degree 1, i.e.\! they are defined over $\bQ$. There are however a few examples of rigid CY 3-folds defined over real quadratic fields\footnote{\ We thank Duco van Straten for pointing out such examples to us.}. All known examples are defined over a subfield of $\R$.  
 In the math literature there is the natural expectation \cite{arithmetics} that a rigid CY 3-fold is defined over $\bK$ iff
 the elliptic curve of period $\tau$ is defined over $\bK$ i.e.\! if $j(\tau)\in\bK\subset\R$.
 Thus, in all known examples\footnote{\ We thank Noriko Yui for confirming that this is the state of the art.} 
$$
 j(\tau)\in \R.\eqno{(6)}
$$
$j(\tau)$ for rigid Calabi-Yau $X$ has been computed analytically when $X$ has complex-multiplication
\cite{shoen,thesis} and numerically for some other classes of examples \cite{prep,octic}. 

Comparing eqns.(3) and (6) we see that all known rigid Calabi-Yau compactifications lead to a 4d effective theory which preserves $T$, i.e.\! 
have $\theta=0$ or $\pi$. 

We suggest that eqn.(6) is valid for all rigid CY 3-folds, not just for the known ones. In other words,
$\cs_{\cn=2}$ consists of a finite set of points $j(\tau)\in \bP^1(\R)\subset \bP^1(\C)$. The question
in eqn.(5) has a positive answer in the $\cn=2$ context, that is, (conjecturally) quantum consistency implies $T$ invariance of the graviphoton sector.

A physical motivation for $T$-invariance of $\cn=2$ supergravities belonging to the landscape is as follows: consider a (non-rigid) Calabi-Yau $X$ with a mirror $Y$. Computing from the IIA  side, the $\cn=2$ prepotential $F$ is (in our conventions) $i$ times a series with real (in fact rational) coefficients counting  holomorphic spheres in the CY, so that\footnote{\ Here $\epsilon_0=1$, $\epsilon_{\neq0}=-1$.}
$$
\overline{F(X^I)}=-F(\epsilon_I\overline{X}^I)\qquad\quad \left|\begin{array}{l}\text{in the duality frame defined by the}\\
\text{infinite volume limit of mirror CY}\end{array}\right.\eqno{(7)}
$$
 Note that if we had additional vector multiplets the graviphoton coupling would vary with the vev of the scalars in the vector multiplets.  In such a case depending on the vev of the scalars the $T$-symmetry may or may not be preserved.  Therefore $T$-symmetry is not a requirement of $\cn=2$ quantum gravity in general.   Nevertheless, equation (7) shows that there are always loci on the scalar multiplet vevs where $T$-invariance is preserved (in the vector sector).  
Let us assume that this equation remains true also in the rigid case; however in the absence of the large volume limit of the mirror which yields a canonical frame, we need to replace the equality sign in (7) by equality up to $SL(2,\Z)$.
Using  $F(X^0)=-\tau (X^0)^2$, this gives $\tau\sim -\overline{\tau}$, which is the same as $j(\tau)\in\R$. 
Only for the case with no vector multiplet we claim quantum gravity consistency may require $T$-symmetry to be preserved otherwise we can always choose a vev for scalar multiplets where $T$-symmetry is broken.
So preserving $T$-invariance in general theories could not follow for quantum gravitational consistency in the ${\cal N}=2 $ case.
  
In conclusion, quantum consistency fine-tunes the effective photon coupling $\tau$ to a $T$ symmetric value. 
As already noted, we expect that $\tau$ is the expectation value of some fields in a quantum gravitational setup.  For example in the context of type IIB, we can view that $\tau$ as the vev of a field corresponding to the smallest eigenvalue of the Laplacian acting on $(2,1)$ forms on the CY 3-fold.  In particular we have a massive axion, whose vev is either $0$ or $\pi$.  In the Peccei-Quinn setup, this would mean that the minimum of the axion potential pick out these two values.  In explicit examples both possibilities, i.e. $\theta =0$ and $\theta =\pi$ are realized by concrete CY 3-folds.  Typically one may have thought that $\theta =0$ is always the minimum but in this case we see the other option $\theta=\pi$ can in some cases have lower energy. Going through the (short) list of examples for which $\tau$ is computed, indeed one finds that in most instances we have $\theta=0$ while $\theta=\pi$ in the very special case that Calabi-Yau has complex multiplication\footnote{\ As already mentioned in footnote \ref{spei}, $\tau=i$, which corresponds to a Calabi-Yau with complex multiplication by $\bQ[i]$, also has $\theta=\pi$ in a suitable duality-frame.}. There are four known examples of rigid CY 3-folds with complex multiplication having $\tau=(1+\sqrt{-d})/2$ and
$$
(d,h^{1,1})= (1,90),\ (3,36),\ (3,84),\ (7,24).
$$ 

\medskip

In the $\cn=2$ case 
consistency requires, in addition to $\theta=0$ or $\pi$, the fine structure constant $\alpha$, which can be viewed as the vev of the massive saxion field, to satisfy the number theoretic condition\footnote{\ Certain number theoretical aspects of $\tau$ in $\cn=2$ supergravity have been discussed in ref.\cite{moore1,moore2}.}
$$
\text{
$j(i/\alpha)\in \hat\bK$ for $\theta=0$}\quad\text{or}\quad\text{$j(i/\alpha-1/2)\in \hat\bK$ for $\theta=\pi$,}
$$
where $\hat\bK$ is a certain finite extension of $\bQ$.
 We stress that $\cs_{\cn=2}$ is a very small subset of $\{\tau\;|\; j(\tau)\in\hat\bK\}$; 
only finitely many values of the fine structure constant $\alpha=1/\mathrm{Im}\,\tau$ are expected to be quantum consistent.
\medskip

If we lived in a world with $\cn=2$ supergravity coupled to hypermultiplets and no vector multiplets, and if we measured $\theta$-angle for the graviphoton we would find that it is $0$ or $\pi$ with no good explanation based on effective field theory or semiclassical gravity arguments.  However consistency of quantum gravity would explain the frozen values of $\theta$ which thus resolves the `theta-problem' in the context of this hypothetical universe.  Even if our conjecture that only $\theta=0,\pi$ are allowed in the rigid case is false and that there are other values of $\theta$ found in a consistent rigid $\cn =2$ quantum gravity setup, we still have a strange feature and a possible solution to the $\theta$-problem:  the fact that among the finite number of possibilities for the values of $\tau$ there is even one case (let alone most of them) with $\theta=0$ or $\pi$  out of an otherwise arbitrary real parameter set would be hard to explain from the effective field theory perspective.  Moreover, in such a case, it would still lead to a finite probability for the observed value of $\theta$ to be $0$ or $\pi$ because the full allowed set is finite (and small) thus giving a probabilistic solution to the $\theta$-problem.

It is natural to speculate whether this observation can also apply to our universe:  Could it be that a solution to the $\theta$ problem is based on quantum gravity consistency also in our universe?
and the QCD  $\theta$-angle (and similarly for QED) it is set to 0 as one of two possible consistent options with quantum gravity?
Moreover, could it be that $j(i/\alpha)\in\hat\bK$ (i.e.\! to be a small degree algebraic number)  for the fine structure constant $\alpha\sim1/137$?  Unfortunately it does not appear possible to experimentally check this latter statement.

\section*{Acknowledgments}

We have benefited from discussions with Dave Morrison, Duco van Straten, and Noriko Yui.
We would like to thank the hospitality of SCGP where this work was completed during the 2018 Simons Summer workshop.

The research of CV is supported by NSF grant PHY-1067976.

\end{document}